\documentclass[prd,twocolumn,showpacs,amsmath]{revtex4}

\usepackage{bm}

\newcommand{\vecx}{{\bf x}}
\newcommand{\vecn}{{\bf \hat n}}

\newcommand{\vecsigma}{{\bm \sigma}}
\newcommand{\domega}{\partial \Omega}
\newcommand{\E}{{\mathcal E}}

\begin{document}

\title{(Non)existence of static scalar field configurations in finite systems}

\author{Artur B. Adib}
\email{artur.adib@brown.edu}
\affiliation{
  Department of Physics and Astronomy, Dartmouth College,
  Hanover, NH 03755, USA
}
\altaffiliation[Address after September 2002:]{
  Department of Physics, Brown University,
  Providence, RI 02912, USA.
}

\date{\today}

\begin{abstract}
Derrick's theorem on the nonexistence of stable time-independent scalar field configurations
[G. H. Derrick, J. Math. Phys. {\bf 5}, 1252 (1964)] is generalized to {\em finite} systems of
arbitrary dimension. It is shown that the ``dilation'' argument underlying the theorem
hinges upon the fulfillment of specific Neumann boundary conditions, providing
thus new means of evading it without resorting to time-dependence or additional fields of higher spin.
The theorem in its original form is only recovered when the boundary conditions are such that {\em both} the
gradient and potential energies vanish at the boundaries, in which case it establishes the nonexistence
of stable time-independent solutions in finite systems of more than two spatial dimensions.
\end{abstract}


\maketitle


\section{Introduction}
  \label{intro}

It is a remarkable consequence of nonlinearity that certain classical field theories allow for the existence
of localized nondispersive solutions - the so-called solitons or ``classical lumps.'' Apart from generating
a whole subject of its own (see e.g. \cite{dodd}), such solutions provide the starting point for the
computation of important nonperturbative phenomena in a wide class of systems described by path integrals.
To be specific, in the context of quantum (statistical) field theory the relevant framework is the
semiclassical (stationary phase) approximation, which allows one to systematically compute quantum
(thermal) corrections to the classical solutions by expanding the action (partition) functional
around them, engendering unique insights such as the false vacuum decay mechanism or the spectrum of
quantum bound states in particle physics \cite{rajaraman,coleman-book} and the nucleation theory
of Langer in statistical field theory \cite{schulman}. Obtaining these solutions is therefore
the first and one of the most important steps in either case.

For scalar field theories, a celebrated theorem due to Derrick \cite{derrick} (see also
\cite{dodd,rajaraman,coleman-book}) ruled out the existence of stable time-independent
configurations in $D>2$ spatial dimensions, causing a noticeable change in the direction of research
among the practitioners of field theory. This is evident, for example, through the increasing interest in
time-dependent solutions (e.g., nontopological solitons \cite{lee}, $Q$-balls \cite{coleman-qball}
and ``oscillons'' \cite{osc}) or in more complicated models that go beyond the ones considered by
Derrick \cite{rajaraman} (see also \cite{perivola} for more recent proposals).
Similar nonexistence theorems were also established for time-dependent periodic solutions
\cite{pagels}, constrained systems (e.g., the nonlinear sigma model) \cite{weder1} and, more generally,
to a class of relativistic equations satisfying relatively mild conditions \cite{weder2}.

Since the traditional physical arena for the practice of field theory is an unbounded space-time,
the question of how and when possible ``finite-size effects'' are important is a most relevant one in
many circumstances.
For instance, it is a well known and appreciated phenomenon that quantum fields, when subject to a finite
spatial domain, induce a measurable force among its confining boundaries - the so-called Casimir effect,
a phenomenon of growing interest both theoretically and experimentally \cite{casimir-review}.
Other finite-size effects considered in the literature include scenarios such as
interacting scalar fields \cite{carrillo}, fields subject to integrable boundary conditions \cite{bajnok}
or kink-bearing systems \cite{klassen}. More recently, the successful theory of metastable decay
of Coleman and Langer was systematically extended to finite systems \cite{maier-stein} and,
being a semiclassical (stationary phase) theory of scalar fields, the existence of nontrivial
classical solutions is fundamental for its formalism.

Motivated not only by the above applications but also by the lack of results along these lines, in
this paper the effects of a finite volume of arbitrary geometry on the existence and stability of
static scalar field solutions in $D$ spatial dimensions will be considered, generalizing the results of
Ref. \cite{derrick} to confined systems. It will be seen that the finitude of
the system introduces important surface terms that allow one to evade Derrick's theorem by suitably choosing
the kind of boundary conditions (Secs. \ref{hamilt} and \ref{exist-stab}). In some particular cases, however,
Derrick's theorem is recovered in its original form imposing constraints on the existence of static solutions
even in finite systems (Sec. \ref{exist-stab}).


\section{Hamilton's principle {\em vs.} spatial dilation} \label{hamilt}
In order to understand the argument underlying Derrick's theorem and the role of the boundary conditions
in a transparent way, it is worth briefly recalling the usual principle of Hamilton. Consider a multicomponent
real scalar field $\phi_i(\vecx,t)$, with $i=1,\ldots,N$ (e.g., for $N=2n$ it can describe the real components of
$n$ complex fields), in a volume $\Omega$ bounded by a closed surface $\domega$. Hamilton's principle
in its original form is based on an {\em interior} variation of the action, i.e.,
\begin{equation} \label{phi_epsilon}
  \phi_i(\vecx,t) \to \phi_i'(\vecx,t) = \phi_i(\vecx,t) + \epsilon \, \eta_i(\vecx,t),
\end{equation}
where $\epsilon$ controls the amplitude of the perturbation and $\eta_i(\vecx,t)$ are arbitrarily shaped
functions that {\em vanish at the boundaries} $\domega$ (and, of course, at the initial
and final times). With such assumption, the stationarity condition of the action, $\delta S[\phi]=0$, yields
immediately the Euler-Lagrange equations of motion, since the additional surface terms vanish
identically due the condition imposed on $\eta_i(\vecx,t)$.

Derrick's arguments, on the other hand, are based on a {\em global} one-parameter dilation of the
field, namely
\begin{equation} \label{phi_lambda}
  \phi_i(\vecx) \to \phi'_i(\vecx) = \phi_i(\lambda \vecx),
\end{equation}
where now $\lambda$ controls the strength of the dilation and the time-dependence was omitted (we are looking
for static solutions).
In contrast to Eq.~(\ref{phi_epsilon}), this variation takes place everywhere and, since it merely amounts to
``stretching'' or ``shrinking'' the field configuration, one would expect from intuitive grounds that it
would violate immediately whatever boundary conditions the fields are subject to, being therefore an invalid
formulation of Hamilton's principle.
Below it will be shown that this is only partially true, as there are certain boundary conditions
that are actually compatible with the dilation argument, though this also shows that the theorem is
severely restricted by the conditions imposed on each particular problem.
It is also worth mentioning that, since only one degree of freedom was introduced, such variation
cannot encompass the infinite number of functional forms in Hamilton's original principle (thanks to
$\eta_i(\vecx)$), and hence one solution that is
stationary with respect to Eq.~(\ref{phi_lambda}) is not necessarily stationary with respect to
the more general case of Eq.~(\ref{phi_epsilon}), from which one concludes that {\em the theorem can provide
only necessary but never sufficient conditions for the existence and stability of static solutions}.

Given the time-independence of the problem, the energy of the field,
\begin{equation}
  E[\phi_i] = \int_\Omega d^D x \left[ \frac{1}{2} (\nabla\phi_i)\cdot(\nabla\phi_i) + V(\phi_i) \right],
\end{equation}
(henceforth ``$\cdot$'' indicates an internal product over the spatial components and the summation
convention will be assumed for $i$), is identical to $-S[\phi]/T$ (where $T$ is the time interval), and thus the
variational principle can be obtained by setting $\partial E[\phi_i']/\partial\lambda=0$ at $\lambda=1$.
For notational convenience, let
\begin{equation}
  \E_1(\nabla \phi_i)=\frac{1}{2}(\nabla\phi_i)\cdot(\nabla\phi_i) \quad \text{and} \quad \E_2(\phi_i)= V(\phi_i)
\end{equation}
be the gradient and potential energy densities, respectively. Then
\begin{align}
  \left. \frac{\partial E[\phi_i']}{\partial \lambda} \right|_{\lambda=1} & = \int_\Omega d^D x
        \left[ \frac{\partial \E_1}{\partial (\nabla\phi_i)}\cdot\frac{\partial}{\partial\lambda}(\nabla\phi_i') +
               \frac{\partial \E_2}{\partial\phi_i}\frac{\partial\phi_i'}{\partial\lambda} \right]\nonumber \\
     = \int_\Omega d^D x & \left[\nabla\phi_i\cdot\nabla(\nabla\phi_i\cdot\vecx) + \frac{\partial V}{\partial \phi_i}(\nabla\phi_i\cdot\vecx)\right].
\end{align}
Using now the identity
\begin{equation}
  \nabla\phi_i\cdot\nabla(\nabla\phi_i\cdot\vecx) = \nabla \cdot \left[ \nabla\phi_i\,(\nabla\phi_i\cdot\vecx)\right] -
    \nabla^2\phi_i(\nabla\phi_i\cdot\vecx)
\end{equation}
and the divergence theorem, one obtains the global stationarity condition
\begin{align}
\int_\Omega d^D x \left[ - \nabla^2 \phi_i + \frac{\partial V}{\partial \phi_i} \right]&(\nabla \phi_i \cdot \vecx) \nonumber \\
     + \int_{\domega} d^{D-1} &\vecsigma \cdot \nabla \phi_i \, (\nabla \phi_i \cdot \vecx) = 0. \label{first_deriv}
\end{align}
If $\phi_i(\vecx)$ is a static solution of the Euler-Lagrange equations the first integral above
vanishes. The dilation argument then tells us the condition
\begin{equation} \label{first_cond}
  \int_{\domega} d^{D-1}\vecsigma \cdot \nabla \phi_i \, (\nabla \phi_i \cdot \vecx) = 0
\end{equation}
determines the stationarity of the solution. This is clearly not compatible with Hamilton's principle, for
we know that the Euler-Lagrange equations alone are necessary {\em and} sufficient to determine the stationarity
(thence the existence) of the solution \cite{lanczos}. In other words, Eq. (\ref{first_deriv}) is not
a necessary condition for the existence of solutions unless Eq.~(\ref{first_cond}) is {\em a priori}
satisfied (say, by imposing vanishing Neumann boundary conditions), and it is
only in this case that the theorem can provide valid constraints on the existence of static solutions
\footnote{In Ref. \cite{requardt} a similar argument was considered to establish rigorously the validity of
Derrick's theorem for infinite systems. That study showed, in particular, that when augmented with the
additional condition of finite energy, surface terms like the one in Eq.~(\ref{first_cond}) vanish in
the infinite-volume limit, proving thus the validity of the theorem for unbounded systems.}.
This can also be restated as follows:

\begin{itemize}
\item {\em In Neumann problems (i.e., $\vecn\cdot\nabla\phi_i$ specified at
$\domega$), Derrick's arguments are valid if and only if the boundary conditions are such that
Eq.~(\ref{first_cond}) is satisfied (e.g., $\vecn\cdot\nabla\phi_i|_{\domega}=0$)}.

\item {\em In Dirichlet problems ($\phi_i$ specified at $\domega$), Eq. (\ref{first_deriv}) is no longer
a necessary condition for the existence of solutions and thus Derrick's arguments do not apply}.
\end{itemize}

\section{Existence and stability conditions} \label{exist-stab}

Assuming that Eq.~(\ref{first_cond}) is satisfied by the boundary conditions, the central results
of Derrick's paper will now be generalized to finite systems. The theorem rests on the
scaling properties of $\E_1$ and $\E_2$ in order to obtain an alternative expression to Eq.~(\ref{first_deriv}).
It is not difficult to see that, after the change of variables $\vecx\to\vecx'=\lambda\vecx$ under the integrals,
one obtains the scaling relation
\begin{align}
  E[\phi_i'] = \lambda^{2-D} \int_{\Omega'} d^D x \,\E_1(\nabla \phi_i) +
    \lambda^{-D} \int_{\Omega'} d^D x \,\E_2(\phi_i), \label{E_phi_prime}
\end{align}
where the unimportant primes were dropped from the integrands. It should be observed that now we have a
$\lambda$-dependence in the domain of integration itself (indicated
here as $\Omega'$), being therefore important to consider the geometry of
the system explicitly before proceeding to the remaining part of the theorem.

In what follows, I shall consider $D$-dimensional volumes $\Omega$ bounded by spherical surfaces $\domega$
of radius $R$. Similar calculations for other geometries like rectangular boxes or cylinders do not
change the results of this section, as the reader can easily verify, suggesting that the results obtained
here are rather general (see also Appendix). For such geometries, the important relation
\begin{align} \label{partial_int}
  \frac{\partial}{\partial \lambda} \int_{\Omega'} d^D x & \, f(\vecx) =
    \frac{\partial}{\partial \lambda} \int_0^{\lambda R} \! \! dr \int_{\domega} d^{D-1}\sigma(r) \,f(r,\vecsigma) \nonumber \\
   & = \lambda^{D-1} R \int_{\domega} d^{D-1}\sigma(R) \,f(\lambda R,\vecsigma)
\end{align}
is easily obtained.  Derrick's existence condition is derived from the first $\lambda$-derivative
of Eq.~(\ref{E_phi_prime}),
\begin{align} \label{derrick1}
  (2-D)\,E_1 - & D\,E_2 \nonumber \\
    + R \int_{\domega} & d^{D-1}\sigma(R)\left[\E_1(\nabla \phi_i) + \E_2(\phi_i) \right] = 0,
\end{align}
where Eq.~(\ref{partial_int}) was used, $E_1\equiv\int_\Omega d^D x \,\E_1(\nabla \phi_i)$ and
$E_2\equiv\int_\Omega d^D x \,\E_2(\phi_i)$. An immediate conclusion is: {\em Even in problems
where condition (\ref{first_cond}) is fulfilled, the presence of an energy component at the boundaries
of the system provides an additional mechanism for evading Derrick's theorem}, although this now requires
a fine-tuning of the boundary conditions to ensure the vanishing of the above expression.
For example, in the interesting case $D=3$ and $V(\phi_i)\geq 0$ considered by Derrick,
the first two terms above are still negative but now one has the option of choosing
the boundary conditions so that the (positive) surface term cancels them.
It is worth emphasizing, however, that this inevitably introduces a certain amount of energy
at the boundaries of the system
and only the physical problem at hand can tell whether this condition can be realized or not
(for problems with Neumann boundary conditions where Eq.~(\ref{first_cond}) is not satisfied, one has the
additional freedom of choosing $\phi_i|_{\domega}$ such that
$[\E_1(\nabla\phi_i)+\E_2(\phi_i)]_{\domega}=0$ for models allowing $V(\phi_i)<0$ somewhere, so in this
case it {\em is} possible to evade the theorem without introducing energy at the boundaries). Note also that,
for configurations
whose energy density vanishes sufficiently fast at spatial infinity, the above expression reduces to the
usual one (cf. \cite{rajaraman,coleman-book}).

The second variation of the energy will tell us about the stability of the configuration. Performing
it in Eq.~(\ref{E_phi_prime}) one obtains, after some algebra,
\begin{align} \label{derrick2}
  \left.\frac{\partial^2 E[\phi']}{\partial \lambda^2}\right|_{\lambda=1} & = (D-2)(D-1)E_1+D(D+1)E_2 \nonumber \\
    - R \int_{\domega} d^{D-1}\sigma&(R) \left[ (D-3) \E_1(\nabla \phi_i) + (D+1) \E_2(\phi_i) \right] \nonumber \\
    + R \int_{\domega} d^{D-1}\sigma&(R) \left[ \nabla\phi_i\cdot\nabla(\nabla\phi_i\cdot\vecx) +
                                  \frac{\partial V}{\partial \phi_i} (\nabla\phi_i\cdot\vecx) \right].
\end{align}
It seems hopeless to extract any information from the above expression without any additional
information from the boundary conditions. Here I only mention the important case where
$\E_1(\nabla\phi_i)=\E_2(\phi_i)=0$ at $\domega$, so that $|\nabla\phi_i|_{\domega}=0$ in agreement with
Eq.~(\ref{first_cond}) and $(2-D)E_1=D\,E_2$ [from Eq.~(\ref{derrick1})], recovering the usual expression,
\begin{equation} \label{derrick2b}
  \left.\frac{\partial^2 E[\phi']}{\partial \lambda^2}\right|_{\lambda=1} = -2(D-2)E_1.
\end{equation}
An important conclusion can now be drawn from the foregoing analysis: {\em For finite systems
whose boundary conditions imply the vanishing of both the gradient and potential energies at $\domega$, the
existence and stability conditions of Derrick remain valid and general, regardless of the specific size
or geometry of the system}. In this case, all the observations based on Derrick's arguments apply
to finite systems as well, from which the most interesting one seems to be that: {\em There are no stable
time-independent solutions in finite systems of more than two spatial dimensions subject to vanishing
gradient and potential energies at the boundaries}.

\section{Conclusions} \label{concl}

A careful investigation of the ``dilation'' argument on the existence and stability of
time-independent scalar field configurations was presented, generalizing previous results to finite
systems of arbitrary dimension and most geometries of interest. The argument itself was found to be,
in general, incompatible with
Hamilton's principle for finite volumes, as summarized by the validity condition (\ref{first_cond}).
This shows in particular that Dirichlet problems are not amenable to an analysis based on the dilation
argument.
Nonetheless, once such condition is satisfied {\em a priori} (e.g., in Neumann problems
where $\vecn\cdot\nabla\phi_i=0$ at $\domega$), the argument
turns into a theorem that provides valuable existence and stability conditions for finite systems.
For such circumstances, alternative means of evading the theorem were proposed that rely on the presence of
energy at the boundaries of the system. Derrick's theorem in its original form is recovered only when
both the gradient and potential energies vanish at the boundaries.

Apart from the obvious generalization to field theories of higher spin or to time-dependent
configurations, it would also be interesting to use the foregoing observations as guidelines to obtain
explicit static solutions. For example, in Ref. \cite{hale} a numeric method based on the stochastic
minimization of the energy functional was developed to obtain nontrivial static solutions to a wide
class of field theories. In this case, the present results can feed the algorithm with the right
boundary conditions that are able to generate the solutions.

\acknowledgments

It is my pleasure to thank Marcelo Gleiser, Robert Caldwell and Rafael Howell for the
interest and discussions. Special thanks go to Renan Landim for pointing out important typos in
the original manuscript. Financial support was provided by Dartmouth College.

\appendix

\section*{Appendix}
Here a generalization of the results of Sec. \ref{exist-stab} for arbitrary smooth geometries will be
sketched. The approach follows closely that of Ref. \cite{khinchin} for phase space
volumes defined by constant-energy surfaces.
Consider a $(D-1)$-dimensional smooth surface $\domega$ defined by $f(\vecx)=F$, with $F$ a
real constant measuring the length scale of the system (e.g., the radius of a $D$-sphere), such
that $\nabla f$ exists and does not vanish within $\Omega$. Then $\vecn=\nabla f/|\nabla f|$ defines a unit
vector normal to $\domega$. Choose a coordinate system $(f,\vecsigma)$ such that $\vecsigma$ are local
coordinates at $\domega$ (e.g., a set of angle variables). Let $d^{D-1}\sigma(f)$
be an area element of $\domega$ and $dn\equiv\vecn\cdot d\vecx$, then
$d^D x = dn\,d^{D-1}\sigma(f) = df\,d^{D-1}\sigma(f)/|\nabla f|$, whereby
\begin{align}
  \frac{\partial}{\partial \lambda} \int_{\Omega'} d^D x & =
    \frac{\partial}{\partial \lambda} \int_0^{\lambda F} df \int_{\domega} \frac{d^{D-1}\sigma(f)}{|\nabla f|} \nonumber \\
   = F \int_{\domega} &\frac{d^{D-1}\sigma(\lambda F)}{|\nabla f|} = \lambda^{D-1} F \int_{\domega} \frac{d^{D-1}\sigma(F)}{|\nabla f|},
\end{align}
generalizing Eq. (\ref{partial_int}) to arbitrary smooth geometries (note that for a spherical geometry
one has $|\nabla f| = 1$ everywhere). From the above relation the results of Sec. \ref{exist-stab} follow
analogously. With certain care it should apply even to arbitrary sharp-edged volumes, since
in practice one can approximate nonsmooth surfaces with smooth ones as accurately as one desires
(e.g., the surface $f(\vecx)=[(x_1/l_1)^n + (x_2/l_2)^n + \ldots]^{1/n}=1$ can be made as close to
a $D$-dimensional rectangular box of edges $l_1,l_2,\ldots$ as one desires by increasing $n$).


\end{document}